\documentclass[iop,natbib]{emulateapj}
\usepackage[]{natbib}
\usepackage{wrapfig}
\usepackage{amsmath}
\usepackage{apjfonts}

\slugcomment{Unpublished}
\shorttitle{32 Ori Group}
\shortauthors{Shvonski et al.}

\begin{document}

\title{A \textit{Spitzer Space Telescope} Survey for Dusty Debris
  Disks\\ in the Nearby 32 Orionis Group}

\author{
Alexander J. Shvonski$^{1,2}$,
Eric E. Mamajek$^{1,3}$,
Jinyoung Serena Kim$^{4}$,
Michael R. Meyer$^{5}$, and
Mark J. Pecaut$^{6}$}
\affil{$^{1}$Department of Physics \& Astronomy, University of
  Rochester, Rochester, NY, 14627-0171, USA}
\affil{$^{2}$Current address: Department of Physics, Boston College,
  140 Commonwealth Ave., Chestnut Hill, MA, 02467, USA}
\affil{$^{3}$Current address: Jet Propulsion Laboratory, California
  Institute of Technology, M/S 321-100, 4800 Oak Grove Dr., Pasadena,
  CA 91109, USA}
\affil{$^{4}$Steward Observatory, University of Arizona, Tucson, AZ,
  85721, USA}
\affil{$^{5}$University of Michigan, Department of Astronomy, 1085
  S. University, Ann Arbor, MI 48109, USA}
\affil{$^{6}$College of Arts and Sciences, Rockhurst University, 1100
  Rockhurst Rd., Kansas City, MO 64110, USA}

\shorttitle{32 Ori Disks}
\shortauthors{Shvonski et al.}

\begin{abstract} 
  We report \textit{Spitzer Space Telescope} IRAC 3.6, 4.5, 5.8 and
  8\,$\mu$m and MIPS 24 and 70\,$\mu$m observations of the 32 Ori
  Group, a recently discovered nearby stellar association situated
  towards northern Orion.  The proximity of the group ($\sim$93 pc)
  has enabled a sensitive search for circumstellar dust around group
  members, and its age ($\sim$20 Myr) corresponds roughly to an epoch
  thought to be important for terrestrial planet formation in our own
  solar system. We quantify infrared excess emission due to
  circumstellar dust among group members, utilizing available optical
  (e.g. Hipparcos, Tycho) and near-IR (2MASS) photometry in addition
  to the \textit{Spitzer} IR photometry.  We report 4 out of the 14
  objects which exhibit 24\,$\mu$m excess emission more than 4$\sigma$
  above the stellar photosphere ($>$20\%) though lacking excess
  emission at shorter wavelengths: HD 35656 (A0Vn), HD 36338 (F4.5),
  RX\,J0520.5+0616 (K3), and HD 35499 (F4).  Two objects (HD 35656 and
  RX\,J0520.0+0612) have 70\,$\mu$m excesses, although the latter
  lacks 24\,$\mu$m excess emission. The 24\,$\mu$m disk fraction of
  this group is 29$^{+14}_{-9}$\%, which is similar to previous
  findings for groups of comparable ages and places 32 Ori as the
  young stellar group with the 2nd most abundant 24$\mu$m excesses
  among groups lacking accreting T Tauri stars (behind only the
  approximately $\beta$ Pic moving group).  We also model the infrared
  excess emission using circumstellar dust disk models, placing
  constraints on disk parameters including $L_{IR}/L_{*}$, $T_{\rm
    disk}$, characteristic grain distance, and emitting area. The
  $L_{IR}/L_{*}$ for all the stars can be reasonably explained by
  steady state disk evolution.
\end{abstract}


\section{Note to Readers}

{\it This report represents the final report for the master's degree
  research of Alex Shvonski in the Department of Physics \& Astronomy,
  University of Rochester. Preliminary results from this survey were
  presented in a AAS poster \citep{Shvonski10}. The first author has
  moved on from astronomical research, and the other coauthors have
  not had the time to bring the analysis and text up to date. Given
  the new astrophysical interest in the new, nearby young stellar
  group associated with 32 Ori, and interest in the stellar and disk
  properties of its membership, the authors have agreed to post this
  unrefereed paper to arXiv in its original form from 2010 (excepting
  comments in italics). Note that a forthcoming paper by Bell, Murphy,
  and Mamajek (submitted to MNRAS) will present and improved age
  estimate for the 32 Ori group, a survey for new low-mass members,
  and analysis of the WISE infrared photometry. Revised estimates for
  the interstellar reddening for the 32 Ori group members are much
  smaller (E($B-V$) $\simeq$ 0.03 mag) than that calculated for the
  cool stars in this 2010 study.  The original study used intrinsic
  colors appropriate for dwarfs, however using an intrinsic color
  sequence appropriate for pre-MS stars \citep{Pecaut13} now results
  in consistently lower reddening values calculated for both the hot
  and cool stellar populations. Also, the age scale for the $<$40
  Myr-old groups plotted in Figure 7 is already obsolete, with the
  ages of some of the groups having been recently revised upward by
  $\sim$50-100\%\, \citep[e.g.][]{Pecaut12, Bell15, Pecaut16}.}\\

\section{Introduction \label{sec:intro}}

It is widely established that circumstellar disks are progenitors of
planetary systems, although the evolution of such systems is less
understood.  A large fraction of stars, thought to be close to unity,
initially contain circumstellar material and then lose their disks
over time, possibly through processes like accretion, planet
formation, photoevaporation, etc. \citep{Meyer07}.  Primordial disks
become gas-depleted on a timescale of $\sim$1-10 Myr \citep{Haisch01,
  Hillenbrand05}, and plots of primordial disk fraction versus age
show an approximately exponential decay with a characteristic
timescale of $\tau\, \simeq\, $2.5 Myr \citep{Mamajek09}.  The
prevalence of gas-depleted (debris) disks suggests that the
collisional grinding of planetesimals is responsible for dust
production, since grain removal processes such as radiation pressure
and Poynting-Robertson drag would not otherwise allow dust to persist
for timescales observed \citep{Backman93}.  Mid-infrared (mid-IR)
excess emission from micron-sized grains in these debris disks is a
successful disk diagnostic that can be used to characterize individual
objects and examine trends including debris disk frequency vs. age
\citep[e.g.][]{Hillenbrand08}.  Of primary importance are studies of
young stellar groups (i.e. clusters and/or associations) of disparate
ages, which provide snapshots of the disk dispersal process from which
we can infer evolutionary trends.\\

Placing our own solar system in context, the timescale between
$\sim$10-50 Myr is a particularly interesting and dynamic epoch of
planetary evolution \citep{Canup04}.  Radionuclide data (Hf-W
chronometry) suggests that the proto-Earth was impacted by a
Mars-sized body which formed the moon at an age of $\sim$25-30 Myr
\citep{Canup04}.  The frequency and characteristics of debris disks at
such ages has potential implications for models of our own solar
system.  However, well-characterized samples of stellar associations
with ages of $<$10-50 Myr are rare within a few hundred pc of the Sun.
This study contributes a snapshot of circumstellar disk evolution
among a new nearby sample of young stars in this age range.\\

We report on a \textit{Spitzer} IRAC and MIPS survey of debris disks
for 14 members of the recently identified nearby young 32 Ori group.
In \S3 we discuss the current evidence for a young group associated
with 32 Ori, first identified by \citet{Mamajek07}.
In \S4 we summarize our observations, including optical spectroscopy,
IRAC, and MIPS data, and additional optical and IR data
(i.e. Hipparcos, Tycho, 2MASS, IRS spectra).
In \S5 we identify disk candidates by examining color-color relations.
In \S6 we model stellar photospheric emission, compute stellar
luminosities, and place group members on an HR diagram to determine
the age of the group.  We then characterize mid-IR excess emission
using dust disk models.
Finally, in \S7 we discuss the disk fraction within the 32 Ori
association and place our results in the broader context of
circumstellar disk evolution, and summarize our findings in \S8.\\

\section{The 32 Ori Group \label{sec:group}}

\citet{Mamajek07} presented evidence for a young stellar group
associated with the B-type star 32 Ori that would appear to constitute
the first northern, young ($<$100 Myr) stellar group discovered within
100 pc of the Sun.  It is a previously unrecognized stellar group
situated towards northern Orion, but at only $\sim$1/5th the distance
of the well-studied $\lambda$ Ori and Ori OB1 star-forming regions
\citep[$\sim$400 pc; $<$10-15 Myr-old,][]{Hernandez06}. With a
distance of $\sim$93 pc and age of $\sim$20 Myr \citep{Mamajek07}, the
32 Ori Group appears to be unrelated to any star-forming regions.  As
the reality of the group, and our assessment of its membership and
age, is critical to our study, we present a more thorough
characterization.\\

\begin{figure}
 \plotone{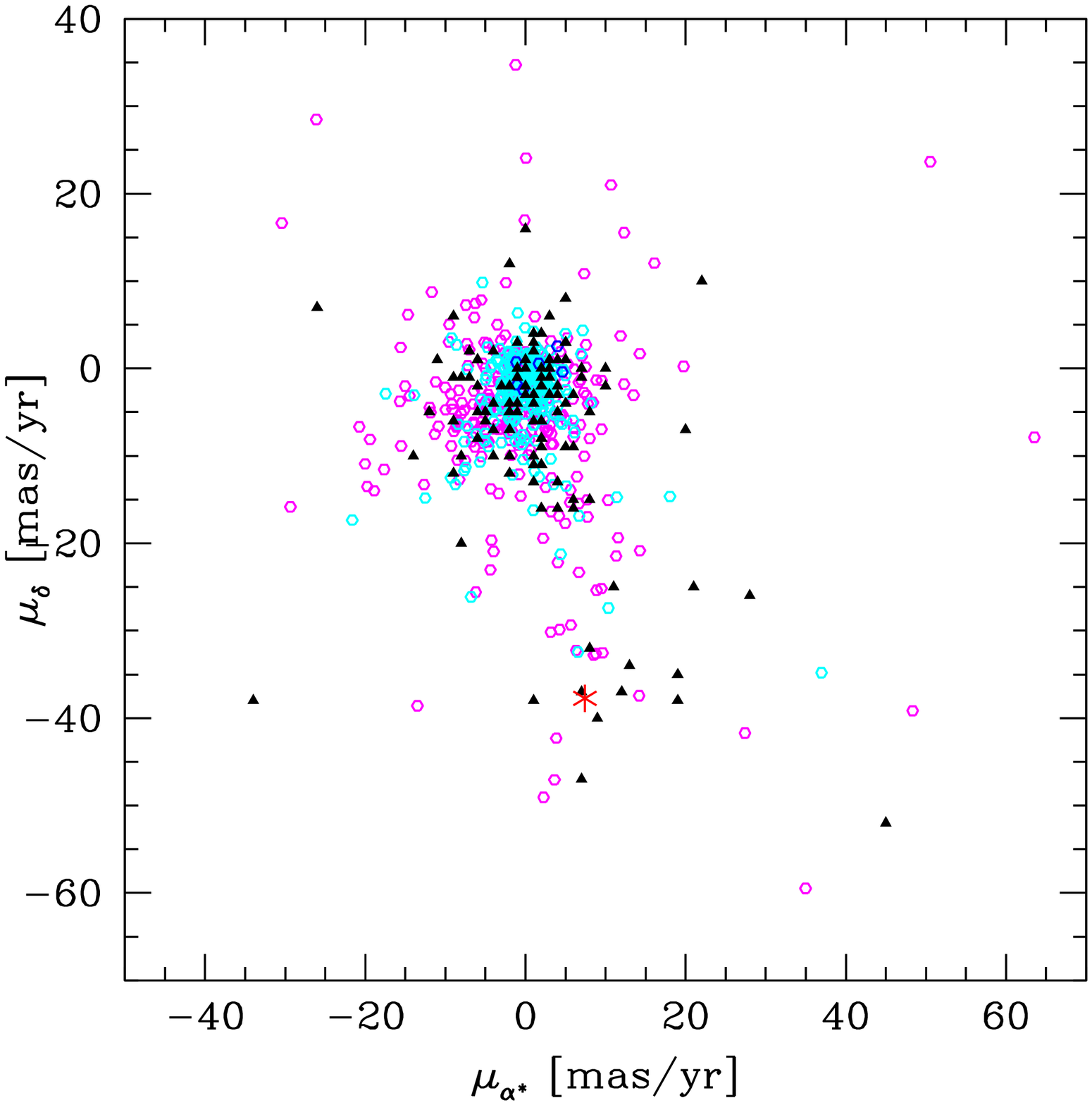}
 \caption{Reproduction of the proper motion plot from
   \citet{Mamajek07} indicating the presense of a young nearby stellar
   group associated with 32 Ori. 32 Ori is marked as a {\it red
     asterisk}. Pre-MS stars within 10$^{\circ}$ of 32 Ori in the
   \citet{Ducourant05} catalog are plotted as {\it filled triangles}.
   Hipparcos stars \citep{ESA97} within 10$^{\circ}$ of 32 Ori of
   O-type ({\it dark blue open circles}), B-type ({\it light blue open
     circles}), and A-type ({\it magenta open circles}).  Note the
   swarm of pre-MS stars with proper motions similar to that of 32
   Ori. The clump of hot stars and pre-MS stars with negligible proper
   motions are associated with Ori OB1 (d $\simeq$ 400 pc).}
\label{pm_discovery}
\end{figure}

The group was originally identifed by \citet{Mamajek07} as a strong
concentration of comoving stars in a vector-point diagram (proper
motion in R.A. vs. proper motion in Dec.) of young stars in northern
Orion (see Fig. \ref{pm_discovery}).  While the more distant $\lambda$
Ori and Ori OB1a members in northern Orion have very small proper
motions ($<$5 mas\,yr$^{-1}$), surrounding the nearby B5+B7 binary 32
Ori \citep[$\varpi$ = 10.77\,$\pm$\,0.64 mas; d = 93$^{+6}_{-5}$ pc;
][]{vanLeeuwen07} is a clump of bright A-stars and X-ray-emitting late
type stars (candidate young stars) with proper motions consistent with
($\mu_{\alpha}$, $\mu_{\delta}$)\, $\simeq$\, (+8, -35)\,
mas\,yr$^{-1}$.  The trigonometric parallaxes of the bright B/A-type
stars in the original \citep{ESA97} and revised \citep{vanLeeuwen07}
are consistent with being approximately co-distant at $\sim$93 pc
(within their parallax uncertainties), although main sequence fitting
suggests more like 100 pc.  Some of the comoving X-ray-luminous stars
were previously discovered to be Li-rich in spectroscopic follow-up of
ROSAT All-Sky Survey (RASS) X-ray sources by \citet{Alcala96,
  Alcala00}.  A subset of the Alcal\'a et al.  RASS sources are very
Li-rich, and share the same radial velocity as 32 Ori ($\sim$18 km
s$^{-1}$). Accurate proper motions from new astrometric catalogs
\citep[e.g. UCAC;][]{Zacharias04, Zacharias10} show that some of these
RASS stars in Orion are co-moving, and are much closer than the Orion
star-forming region. Recent observations by using the FAST
spectrograph on the Mt. Hopkins 1.5m (60'') telescope have led to the
identification of a few new, pre-MS members (\S4.1).\\

\begin{center}
 \begin{deluxetable}{lr}
  \tabletypesize{\scriptsize}
  \tablewidth{0pt}
  \tablecaption{2MASS Identifiers for 32 Ori Group Candidate Members \label{tab:2MASS}}
  \tablehead{
  \colhead{ID} &
  \colhead{2MASS}
}
  \startdata
32 Ori          & J05304706+0556536 \\
2UCAC 33699961  & J05194398+0535021 \\
2UCAC 33878694  & J05253253+0625336 \\
HBC 443         & J05343491+1007062 \\
HD 35499        & J05251457+0411482 \\
HD 35656        & J05263883+0652071 \\
HD 35695        & J05265202+0628227 \\
HD 35714        & J05265999+0710131 \\
HD 36338        & J05311570+0539461 \\
RX\,J0520.0+0612 & J05200029+0613036 \\
RX\,J0520.5+0616 & J05203182+0616115 \\
RX\,J0523.7+0652 & J05234246+0651581 \\
RX\,J0529.3+1210 & J05291899+1209295 \\
HD 245567       & J05371844+1334525
\enddata
%
%
\end{deluxetable}
\end{center}

\

\section{Observations \& Data Reduction \label{sec:obs}}

\subsection{Optical Spectroscopy \label{sec:FAST}}

Optical spectra of the stars listed in Table \ref{tab:2MASS} were
obtained with the FAST instrument on the Tillinghast 1.5-m telescope
at the Fred Lawrence Whipple Observatory in queue mode on UT 25 and 27
Jan 2006.  The spectra cover $\sim3500-7400$\,\AA\, with a resolution
of $\sim3$\,\AA\, and were visually examined using the \textit{sptool}
spectral comparison program described in \citet{Pecaut12} to obtain
spectral types.  A dense grid of MK spectral standards \citep[and
  references therein][]{Gray09}\footnote{Extensive notes on the
  pedigrees of all known dwarf, subgiant, and giant MK standard stars
  can be found at: http://www.pas.rochester.edu/$~$emamajek/spt/} was
assembled using spectra taken at similar resolution from the CTIO
1.5-m telescope and Nstars
website\footnote{http://stellar.phys.appstate.edu}.\\

Several cool stars exhibit H$\alpha$ emission indicative of enhanced
chromospheric activity.  One M-type member, 2UCAC 33699961, has
H$\alpha$ emission that is relatively high (EW$\sim 10.5$\,\AA) but is
still consistent with chromospheric activity rather than accretion
\citep{Barrado03}.  We also measured Li equivalent widths for each
object from the \ion{Li}{1} $\lambda6707$\,\AA\, feature as a youth
indicator. Red optical Spectra and equivalent widths for H$\alpha$ and
\ion{Li}{1} $\lambda6707$\,\AA\, for the K/M-type 32 Ori members are
presented in Figure 1.  Equivalent widths were measured in
IRAF\footnote{IRAF is distributed by the National Optical Astronomy
  Observatory, which is operated by the Association of Universities
  for Research in Astronomy, Inc., under cooperative agreement with
  the National Science Foundation.}. The two M3-type stars show very
strong chromospheric activity (H$\alpha$ emission) but depleted Li.
This suggests that the stars are young, but probably older than 10$^7$
yr \cite[older than $\eta$ Cha and TW Hya groups;][]{Mentuch08}.\\

\subsection{IRAC}

\subsubsection{Observations}

\textit{Spitzer} IRAC observations were obtained 16-17 October 2007 at
3.6, 4.5, 5.8 and 8\,$\mu$m in stellar mode for 11 objects.
Additional data for HBC 443 and HD 245567 was retrieved from the
\textit{Spitzer} archive (program ID 37, PI Giovanni Fazio; program ID
148, PI Michael R. Meyer, respectively).  Observations of HBC 443 were
taken in high dynamic range (HDR) mode on 12 October 2004, while
observations of HD 245567 were taken in subarray mode 8 October 2004.
Observations of HD 245567 were not taken at 5.8\,$\mu$$m$.  IRAC
observations for 32 Ori were not taken, however, we obtained IRS data
from the literature, which will be discussed in \S4.5.  IRAC subarrays
are 32$\times$32 pixel sections of the full array, which is
256$\times$256 pixles in size with a pixel scale of $\sim$1.2\arcsec\,
pix$^{-1}$ (IRAC Data Handbook 3.0\footnote{See
  http://ssc.spitzer.caltech.edu/irac/iracinstrumenthand-book/}).\\

\begin{figure}
 \epsscale{1.2}
 \plotone{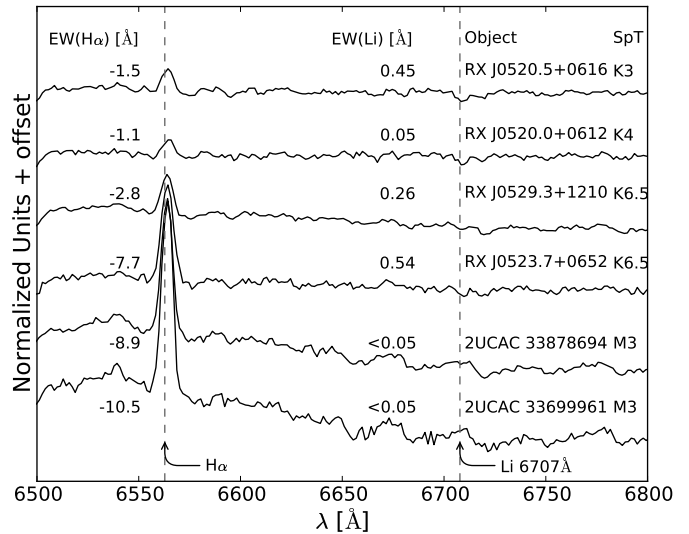}
 \caption{Optical spectra of late-type ($>$K0) group members.
   H$\alpha$ emission indicative of enhanced chromospheric activity is
   clearly apparent.}
\label{hr_diagram}
\end{figure}


The primary reason for the IRAC observations is to complement the
2MASS photometry to predict stellar photospheric emission at 24 and
70\,$\mu$$m$. There is a chance of an excess due to warm dust in the
IRAC bands, however this is very unlikely given that the group is
$>$10 Myr old \citep[e.g.][]{Mamajek04}.  We aimed to detect the
stellar photosphere at a minimum S/N of 50, thereby ensuring similar
relative accuracy as the 2MASS JHK fluxes.  Exposure times in stellar
mode were 0.4 sec for IRAC channels 1 and 2 and 2 sec for channels 3
and 4.  Using a 9 position random dither pattern, this resulted in
integration times of 3.6 sec and 18 sec for the respective channels.\\


\subsubsection{Image Processing}

For all 13 objects, IRAC images were processed with the
\textit{Spitzer} Science Center (SSC) pipeline version S18.7.0.  This
pipeline produces basic calibrated data (BCD) images, for which
reduction steps including dark subtraction and a flat-field correction
are carried out (IRAC Data Handbook 3.0).  The ``mux-bleed'' effect is
also corrected in stellar and HDR mode, but not in subarray mode.
Additionally, nonlinear pixel responses and cosmic ray detections are
flagged in an image mask (IRAC Data Handbook 3.0).\\

BCD images were processed using the SSC package MOPEX
\citep{Makovoz05}.  Briefly, we discuss the data reduction steps
implemented using this package.\footnote{The following steps were
  adopted from the MOPEX User's Guide 18.3.2, which can be found at
  http://ssc.spitzer.caltech.edu/dataanalysistools/tools/mopex/
  mopexusersguide/} For each object, background matching between
overlapping input frames was performed by computing additive offsets
for each image.  First, the median value in a 45$\times$45 pixel
window was subtracted at each pixel location in every image in order
to assist in the detection of bright sources, which were not included
in background matching estimates.  Bright sources were detected in
background subtracted images through image segmentation, and these
sources were masked during background matching.  Background subtracted
images were projected onto a Fiducial Image Frame (FIF), which defines
the overlap between all images.  Ultimately, offsets were computed by
minimizing the differences between overlapping pixels and requiring
that the sum of the offsets equal zero.  Offsets that were 3$\sigma$
outliers were not included in computing the offsets for other images.\\

Before photometry was performed on background matched images, they
were used to create a set of outlier masks.  Multiframe outlier
detection was performed by filtering individual pixels at common
spatial locations in background corrected images for time dependent
phenomena.  The median absolute deviation of individual pixels was
computed and outliers deviating more than 5 times this value from the
median were rejected.  The results of multiframe outlier detection
were used to create an ``rmask.''  Pixels masked in the rmask were
reinterpolated in the background matched images.  In addition to
rmasks, pmasks (containing permanently damaged pixels), and DCE (data
collection event) status masks (containing temporarily damaged pixels)
are also used to mask pixels.\\

Additional care must be taken for the reduction of IRAC images.  The
position of the IRAC instrument causes a variation in pixel solid
angle of order $\sim$1.5\% across each channel (IRAC Data Handbook
3.0).  Additionally, the effective filter bandpass varies across the
array of each IRAC channel, producing differences in flux of order
$\sim$10\% \citep[IRAC Data Handbook 3.0;][]{Quijada04}.  Thus, there
are two important corrections that must be performed on BCD data: a
pixel solid angle correction (PSAC)\footnote{Correction images are
  available at http://ssc.spitzer.caltech
  .edu/irac/calibrationfiles/solidangles/} and an
array-location-dependent optical distortion correction
(ALDODC)\footnote{Correction images are available at
  http://ssc.spitzer.caltech
  .edu/irac/calibrationfiles/locationcolor/}.\\

Corrections that account for optical distortion and pixel solid angle
variations across IRAC arrays are encoded in the headers of BCD images
(IRAC Data Handbook 3.0).  The SSC provides ALDODC images, which also
include PSAC information in their headers.  Therefore, we divided the
ALDODC images by the PSAC images before applying this correction, so
that the PSAC was not applied twice.  An important caveat to the above
corrections is that they are only applicable to objects emitting in
the Rayleigh-Jeans regime at IRAC wavelengths (e.g. photospheric
emission from point-sources) (IRAC Data Handbook 3.0).  Since 32 Ori
members, due to their age, most likely do not have warm inner disks,
we expect that their emission will be purely photospheric and no
additional precautions will need to be heeded.\\

\subsubsection{Photometry}

Photometry was performed using the APEX routine aperture.pl, using an
aperture radius of 3 image pixels and a sky annulus of 12-20 image
pixels.  This routine computes aperture photometry for a specified
point in a set of images; it applies pmasks, DCE status masks, and
rmasks, as described in the previous section.  As input, we used the
background matched BCD images corrected for pixel solid angles and
optical distortion.  We used the median fluxes obtained from sets of
corrected images as final values and standard errors as measures of
uncertainty.  Additionally, there exist uncertainties in the
calibration factor used to convert instrumental units to MJy/sr.
Uncertainties for channels 1-4 are 1.8\%, 1.9\%, 2.0\%, and 2.1\%,
respectively \citep{Reach05}.  \citet{Carpenter08} compared the flux
density calibrations used by \citet{Reach05} in full-array mode with
sub-array data to determine if there exist relative offsets in any
IRAC channels.  They determined that differences are of order $<$1\%
and, therefore, inconsequential.  Thus, we perform no additional
correction to data taken in sub-array mode.  Aperture corrections for
all IRAC channels were obtained from the IRAC Data Handbook 3.0; they
are 1.112, 1.113, 1.125, and 1.218 for channels 1, 2, 3, and 4,
respectively.  We used zero-point fluxes reported in the IRAC Data
Handbook 3.0 to convert flux measurements to magnitudes.\\

\subsection{MIPS 24\,$\mu$$m$}

\subsubsection{Observations}

\textit{Spitzer} MIPS observations were obtained 16-17 March 2008 at
24 and 70\,$\mu$m in photometry mode.  A total of 12 objects were
imaged at 24\,$\mu$m and 13 objects (including 32 Ori) were imaged at
70\,$\mu$m.  Additional 24 and 70\,$\mu$m data for one object, HD
245567 (observed 13 October 2004), were obtained from the
\textit{Spitzer} archive (program ID 148, PI Michael R. Meyer).  In
this section, we describe data reduction and photometry performed on
the 24\,$\mu$m data; the 70\,$\mu$m data are discussed in section 4.\\


The MIPS 24\,$\mu$m array is 128x128 pixels in size with a field of
view of $\sim$5.4\arcmin$\times$5.4\arcmin, giving it a scale of
2.49\arcsec$\times$2.60\arcsec\, pix$^{-1}$ (MIPS Data Handbook
3.3.1\footnote{http://ssc.spitzer.caltech.edu/mips/mipsinstrumenthandbook/}).
Our goal was to detect stellar photospheres at minimum S/N of 10,
where the predicted photospheric fluxes range from $\sim$1.00 to 18.65
mJy for our targets. Thus, we performed 3 sec integrations for B
through early-K type stars, using either 2 or 4 cycles, and 10 sec
integrations for later type stars, using up to 18 cycles for an M3
star.\\

\subsubsection{Image Processing}

For 12 objects, MIPS 24\,$\mu$m images were processed with the SSC
pipeline version S17.0.4.  SSC pipeline version S16.1.0 was used to
process HD 245567 data.  Similarly, these pipelines produce basic
calibrated data (BCD) images, which were processed using the SSC
package MOPEX \citep{Makovoz05} in the same manner as the IRAC data.
An additional step was taken to create mosaic images from background
matched BCD images, which were used for point source detection in
APEX.  Mosaic images for each object were created by first applying
rmasks, pmasks, and DCE status masks and then combining images using
average pixel values.  MIPS data does not require additional
corrections from optical distortion and variations in pixel solid
angle; therefore, we did not apply PSAC or ALDODC images.\\


\begin{center}
 \begin{deluxetable*}{lrrrrrrr}
  \tabletypesize{\scriptsize}
  \tablewidth{0pt}
  \tablecaption{{\it Spitzer} IRAC \& MIPS Photometry}
  \tablehead{
  \colhead{ID} &
  \colhead{2MASS} &
  \colhead{[3.6] (mag)} &
  \colhead{[4.5] (mag)} &
  \colhead{[5.8] (mag)} &
  \colhead{[8.0] (mag)} &
  \colhead{[24] (mJy)} &
  \colhead{[70] (mJy)}
}
  \startdata
32 Ori          & J05304706+0556536 & $\cdots$      & $\cdots$      & $\cdots$      & $\cdots$      & $\cdots$      & 9.8$\pm$2.3\\
2UCAC 33699961  & J05194398+0535021 & 9.76$\pm$0.04 & 9.72$\pm$0.07 & 9.70$\pm$0.02 & 9.68$\pm$0.04 & 1.10$\pm$0.01 & -2.8$\pm$1.6\\
2UCAC 33878694  & J05253253+0625336 & 9.56$\pm$0.03 & 9.58$\pm$0.04 & 9.50$\pm$0.03 & 9.50$\pm$0.02 & 1.17$\pm$0.02 & -6.8$\pm$3.0\\
HBC 443         & J05343491+1007062 & 7.36$\pm$0.02 & 7.35$\pm$0.02 & 7.39$\pm$0.09 & 7.33$\pm$0.04 & 8.94$\pm$0.14 & 6.4$\pm$6.0\\
HD 35499        & J05251457+0411482 & 7.48$\pm$0.01 & 7.49$\pm$0.01 & 7.52$\pm$0.01 & 7.48$\pm$0.03 & 8.45$\pm$0.10 & 1.8$\pm$1.3\\
HD 35656        & J05263883+0652071 & 6.48$\pm$0.01 & 6.48$\pm$0.02 & 6.51$\pm$0.01 & 6.47$\pm$0.00 & 88.45$\pm$0.36 & 91.0$\pm$4.2\\
HD 35695        & J05265202+0628227 & 7.67$\pm$0.03 & 7.67$\pm$0.02 & 7.67$\pm$0.01 & 7.64$\pm$0.01 & 6.30$\pm$0.20 & 13.1$\pm$5.9\\
HD 35714        & J05265999+0710131 & 6.75$\pm$0.02 & 6.76$\pm$0.02 & 6.77$\pm$0.01 & 6.78$\pm$0.01 & 13.33$\pm$0.13 & 1.1$\pm$2.3\\
HD 36338        & J05311570+0539461 & 7.41$\pm$0.02 & 7.41$\pm$0.01 & 7.42$\pm$0.00 & 7.41$\pm$0.01 & 14.79$\pm$0.15 & 8.6$\pm$2.9\\
RX\,J0520.0+0612 & J05200029+0613036 & 8.53$\pm$0.02 & 8.55$\pm$0.02 & 8.52$\pm$0.02 & 8.48$\pm$0.01 & 2.91$\pm$0.06 & 7.9$\pm$1.7\\
RX\,J0520.5+0616 & J05203182+0616115 & 8.49$\pm$0.02 & 8.49$\pm$0.03 & 8.45$\pm$0.02 & 8.41$\pm$0.01 & 3.87$\pm$0.09 & -3.5$\pm$1.9\\
RX\,J0523.7+0652 & J05234246+0651581 & 8.94$\pm$0.02 & 8.93$\pm$0.03 & 8.90$\pm$0.02 & 8.87$\pm$0.02 & 1.99$\pm$0.03 & 7.8$\pm$3.1\\
RX\,J0529.3+1210 & J05291899+1209295 & 9.02$\pm$0.02 & 9.03$\pm$0.04 & 9.00$\pm$0.03 & 8.99$\pm$0.01 & 1.63$\pm$0.03 & 11.0$\pm$8.3\\
HD 245567       & J05371844+1334525 & 7.51$\pm$0.02 & 7.52$\pm$0.03 & $\cdots$      & 7.50$\pm$0.09 & 6.87$\pm$0.09 & -24.9$\pm$5.7
\enddata
\tablecomments{``$\cdots$'' represent photometry that we did not
  obtain.  Additional absolute calibration uncertainties of 1.8\%,
  1.9\%, 2.0\%, 2.1\%, 4\%, and 7\% for respective IRAC and MIPS
  channels \citep[MIPS Data Handbook 3.3.1;][]{Reach05,Engelbracht07}
  should be added in quadrature with the photometric uncertainties
  reported here.}
 \end{deluxetable*}
\end{center}

\subsubsection{Photometry}

Multiframe point source extraction was implemented using the SSC APEX
package.  In this process, sources are detected using the mosiac image
while a PRF is simultaneously fit to a stack of inputs images,
ultimately giving a flux density for detected sources.  We utilized
the following procedure.\footnote{see MOPEX User's Guide 18.3.2 for a
  complete discussion of PRF fitting.}  First, background subtraction
of the mosaicked images was performed using a median value computed
from a 25$\times$25 pixel window at each pixel in preparation for
point source detection.  Next, non-linear matched filtering was
performed on the background subtracted images in order to improve the
detectability of point sources.  Point source probability (PSP)
images, which measure the probability that there is a point source
above the noise level at each pixel, were created.  These images were
used as input into the detect module, which performs image
segmentation to detect sources above a 4$\sigma$ threshold.  Similar
to the process of point source \textit{detection}, the stack of input
images was background subtracted in preparation for point source
\textit{estimation} using a median value computed from a 45$\times$45
pixel window at each pixel.  PRF fitting was performed on these images
using a 7$\times$7 pixel fitting area.\\

\citet{Carpenter08} noted that the uncertainties computed by the APEX
module are intrinsically smaller than errors computed from the
repeatability of observations.  Thus, we computed standard errors from
aperture photometry performed on individual BCD images and adopted
these values as a measure of repeatability; aperture photometry was
performed using the APEX routine aperture.pl.  As done with IRAC
images, background matched BCD images were masked using pmasks, status
masks, and rmasks.  For all sources except HD 35656, a smaller
aperture radius of 3.5\arcsec\, and annulus size of
20\arcsec-32\arcsec\, were used, with a corresponding aperture
correction of 2.57.  For HD 35656, a larger aperture radius of
7\arcsec\, and annulus of 20\arcsec-32\arcsec\, were used, with a
corresponding aperture correction of 1.61.  Aperture corrections were
obtained from the MIPS Data Handbook 3.3.1.  Additionally, we adopted
a 24\,$\mu$m calibration uncertainty of 4\%\, \citep{Engelbracht07}.\\

We found that PRF estimates of flux density deviated from aperture
photometry measurements by $<$10\% for all objects except for RX
J0529.3+1210. This star was identified as a spectroscopic binary by
\citet{Mace09}, and was located in a region of complex nebulosity,
which compounded the difficulty in obtaining accurate photometry.

\subsection{MIPS 70\,$\mu$$m$}
\subsubsection{Observations}

We obtained 70\,$\mu$m data in wide photometry mode, where the array
of 32$\times$32 pixels has a scale of
9.85\arcsec$\times$10.06\arcsec\, pix$^{-1}$.  Again, 13 objects were
observed while data for HD 245567 was obtained from the
\textit{Spitzer} archive.\\

We realized that reaching the photosphere for all of our objects,
ranging in spectral type from B-M, at 70\,$\mu$m was unfeasible, so we
instead adopted a uniform survey depth of 18 cycles of 10 sec frame
time for each star, giving approximate 1$\sigma$ point source
sensitivities of $\sim$3 mJy for our background levels. In the absense
of a disk, the brightest star (32 Ori; F$_{70,phot}\sim$10 mJy) would
have its photosphere detected at $\sim$4$\sigma$ with this observing
sequence; however, the photospheres of the other objects
(F$_{70,phot}$ $\sim$0.1-2.1 mJy) are below this 1$\sigma$ limit.  A
disk emitting at $\sim$9000 $Z$ \citep[$Z$ = 1 zody =
  1.125$\times$10$^{21}$ cm$^{2}$;][]{Gaidos99} would be detectable at
3$\sigma$ around our targets, assuming a 70\,$\mu$m Wien temperature.
For comparison, a survey of similar depth in the similarly aged
cluster NGC 2547 \citep[d=450 pc;][]{Young04} could detect only disks
emitting at $>2\times 10^{5}$ $Z$.\\

\subsubsection{Image Processing}

All MIPS 70\,$\mu$m images were processed with the SSC pipeline
version S17.0.4 except HD 245567, for which pipeline version S16.1.0
was used.  This work utilized filtered basic calibrated data (FBCD),
produced by subtracting the temporal median of adjacent DCEs from each
pixel in order to account for latent phenomena (MIPS Data Handbook
3.3.1).  FBCD images were processed using the SSC package MOPEX
\citep{Makovoz05} in the same manner as the 24\,$\mu$m data in order
to create mosaic images.\\



\subsubsection{Photometry}

Aperture photometry was performed on the mosaicked images in IRAF
using an aperture radius of 16\arcsec\, and an annulus size of
18\arcsec-39\arcsec, corresponding to 2$\times$HWHM (MIPS Data
Handbook 3.3.1).  This method is preferable to PRF fitting because
most sources are not detectable in individual FBCD images.  Apertures
were centered on positions determined from the detection procedure in
the 24\,$\mu$m APEX module.  An aperture correction of 2.07 was
adopted from the MIPS Data Handbook 3.3.1, corresponding to a 60 K
blackbody.\\

RX\,J0520.5+0616 appears to have been contaminated by a nearby (likely)
extragalactic source.  We performed PRF fitting and subtraction of the
source and redid the 70\,$\mu$m photometry, as described above.  This
step was not needed at 24\,$\mu$m because of the simultaneous PRF
fitting that was performed.  The nearby object was not visible at
near-IR wavelengths.  No spurious point sources were detected with the
aperture/annulus around other objects.\\

The uncertainties in the 70\,$\mu$m images were computed following the
prescription of \citet{Carpenter08}.  We used the following equation:\\

\begin{equation}
\sigma=(\eta_{sky}\eta_{corr})(\Omega \Sigma_{sky})\sqrt{N_{ap}+N_{ap}^{2}/N_{sky}}
\end{equation} 

The values in equation (1) are the number of pixels in the aperture
$N_{ap}$, the number of pixels in the annulus $N_{sky}$, the solid
angle/pixel $\Omega$, the noise/pixel within the sky annulus
$\Sigma_{sky}$, the correlation correction for adjacent pixels
$\eta_{corr}$, and the correction for systematic differences between
aperture and annulus background noise $\eta_{sky}$.  The first four
values were computed in IRAF.  The correlation correction is necessary
because FBCD images are interpolated in order to produce a final
mosaic that is 2.5 times larger than the initial images (i.e. FBCDs of
9.8\arcsec\, pixel$^{-1}$ are transformed to a scale of 4\arcsec\,
pixel$^{-1}$).  Thus, the correlation correction is $\eta_{corr}$=2.5.
The correction for disparate aperture/annulus noise is necessary
because there exists higher noise in pixel columns near the source
position; \citet{Carpenter08} concluded that this effect was due to
latent phenomena from calibration stim flashes.  We adopt their value
of $\eta_{sky}$\,=\,1.40.  The additional flux calibration uncertainty
of 7\% was used (MIPS Data Handbook 3.3.1).\\

\subsection{Additional Photometry}

We obtained $B$ and $V$-band photometry and photometric errors from
the Hipparcos \citep{ESA97} and Tycho-2 catalogues \citep{Hog00}.  We
converted Tycho-2 photometry to the Johnson/Cousins system using the
polynomial fits from \citet{Mamajek02,Mamajek06}.  For 32 Ori, we
retrieved two additional IR fluxes from \citet{Chen06}, which
correspond to the photometric bands 8.5-13\,$\mu$m and 30-34\,$\mu$m,
computed from IRS spectra.  The fluxes, which are not listed in Table
1, are 468$\pm$0.5 mJy and 50.4$\pm$1.7 mJy, respectively, and have
calibration errors of 5\% \citep{Chen06}.\\

\begin{figure*}
 \epsscale{1} \plotone{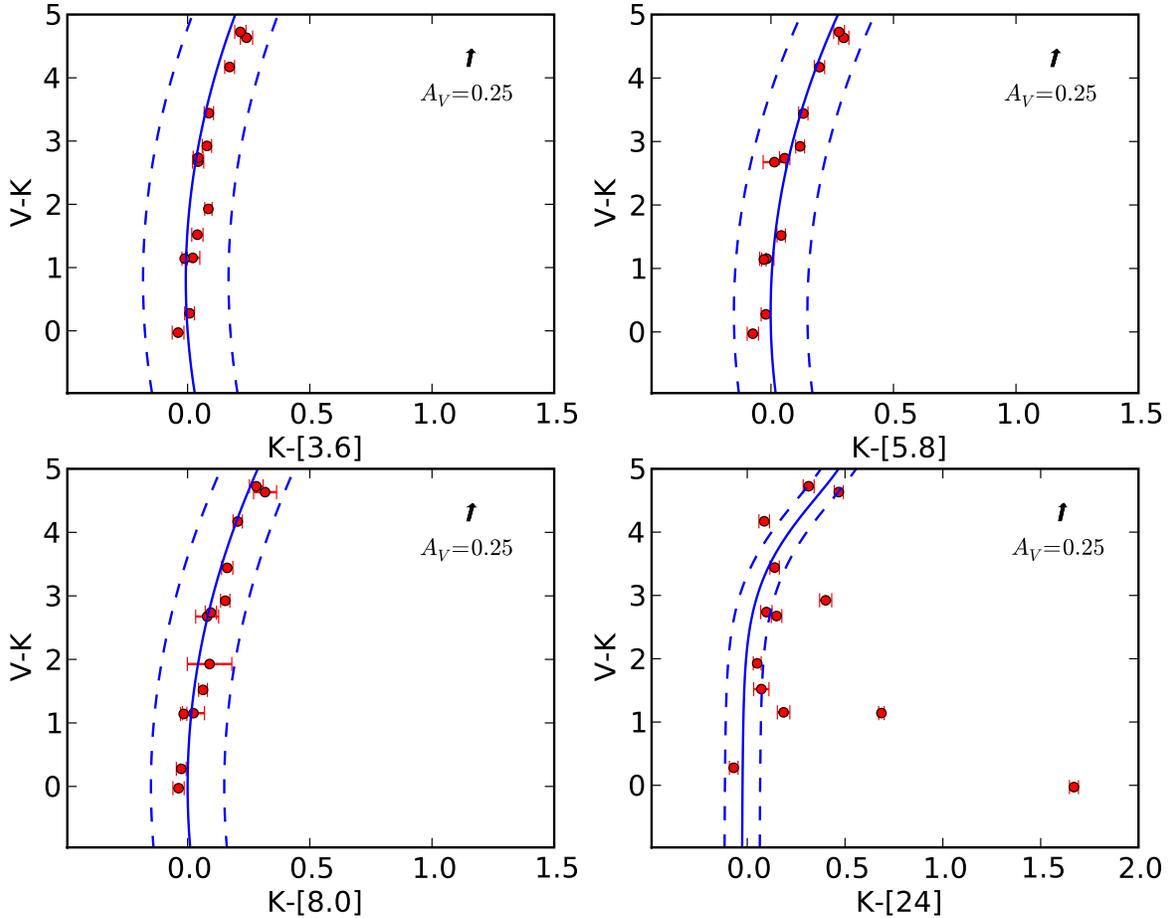}
\caption{$V-K_s$ vs. $K_{s}-$[$\lambda_{\mu m}$] color-color diagrams of 32 Ori
  group members.  The solid line in each plot represent the
  zero-excess locus from \citet{Balog09}; the dashed lines represent
  the 3$\sigma$ rms scatter ($\sim$17.5\%, $\sim$15\%, $\sim$15\%, and
  $\sim$9\%, respectively) .  Horizontal error bars represent
  1$\sigma$ photometric errors.  In each plot, the vector $A_{V}$
  represents the average reddening towards group members.}
\end{figure*}

\section{Identifying Disk Candidates \label{sec:identdisk}}

\subsection{IRAC Excesses}

Longer wavelength emission probes increasingly cooler circumstellar
material at larger radii.  Excess emission at IRAC wavelengths likely
originates from a region close to the star ($<$1AU)
\citep[e.g.][]{Balog09} and is, therefore, a warm inner disk
diagnostic.  Color-color diagrams indicate whether objects exhibit
redder colors than an intrinsic photospheric value given their
spectral type, where $V-K_s$ serves as a stellar temperature
indicator.  3.6, 5.8, and 8\,$\mu$m photospheric excesses for group
members were determined from $V-K_s$ versus $K_{s}-$[$\lambda_{\mu
    m}$] color-color diagrams (Fig. 2).  We adopted intrinsic stellar
color loci of \citet{Balog09} for all of these plots.  The loci were
derived by fitting a second degree polynomial to color-color diagrams
of $\sim$140 members of NGC 2457A and B \citep{Balog09}.\\


Clearly, these figures indicate that no objects exhibit excess
emission above the photosphere at IRAC wavelengths, as no objects
appear statistically significantly redder than their intrinsic stellar
loci.  We anticipate no near-IR excesses due to the age of the 32 Ori
group \citep[e.g.][]{Mamajek04}.  Ultimately, this analysis is
consistent with spectroscopic results and implies that group members
are not accreting T Tauri stars and do not have detectable warm inner
dust disks.\\

\subsection{MIPS Excesses}

Figure 2 shows the $V-K_s$ versus $K_s$-[24] color-color diagram for 32 Ori
members.  We adopted the intrinsic color locus of \citet{Balog09},
derived from 1500 stars from numerous \textit{Spitzer} debris disk
programs.  \citet{Balog09} fit a seventh-order polynomial (eq. 2) to
the color-color locus and determined an rms scatter of $\sim$3\%.\\

\begin{equation*}
 (K_{s} - [24])_{0} = -0.024 + 0.0025 (V-K)_{0} 
\end{equation*}
\begin{equation*}
+ 0.001 (V-K_{s})_{0}^{2} + 0.0005 (V-K_{s})_{0}^{3} 
\end{equation*}
\begin{equation}
+ 0.00012 (V-K_{s})_{0}^{6} - 0.000019 (V-K_{s})_{0}^{7}
\end{equation}

It is apparent from Figure 2 that several 32 Ori members exhibit
24\,$\mu$m photospheric excesses.  We determined that 4 members,
namely HD 35656 (A0Vn), HD 36338 (F4.5), RX\,J0520.5+0616 (K3), and HD
35499 (F4), have excess emission more than 4$\sigma$ ($\geq$20\%)
above their stellar photospheres, indicating that these members are
debris disk candidates.  Additionally, 3 of these 4 members (excluding
HD 35499) have excess emission more than $\geq$38\% above their
photospheres.  Thus, the disk fraction within this group at the
4$\sigma$ excess level is 29\% (+14\%, -9\%; 1$\sigma$ upper/lower
limits, respectively), using binomial statistics
\citep{Burgasser03}.\\

We determined 70\,$\mu$m photospheric excesses by assuming a
Rayleigh-Jeans dependence, which implies that
$F_{70,phot}=F_{24,phot}(24/70)^{2}$.  Thus, the only parameter
required to determine $F_{70,phot}$ was the 24\,$\mu$m photospheric
flux derived from the \citet{Balog09} color-color locus.  We found
that 2 objects had 70\,$\mu$m excess emission $>$4$\sigma$ above their
photospheric values: HD 35656, which also exhibited a 24\,$\mu$m
excess, and RX\,J0520.0+0612, which did not exhibit a 24\,$\mu$m
excess.  70\,$\mu$m emission for all other objects (except 32 Ori) was
statistically zero, implying a non-detection.  For 32 Ori, we detected
emission consistent with predicted photospheric flux.\\

Interestingly, HD 35656 has one of the largest 24\,$\mu$m excesses
among objects of similar ages.  The ratio of observed 24\,$\mu$m flux
to intrinsic photospheric emission, as determined from the K-[24]
locus, is $\sim$4.76.  Subtracting the predicted photospheric flux
from the observed flux indicates that emission from a putative disk is
$\sim$376\% greater than that from the star alone.  \citet{Balog09}
searched the literature and determined that only 8 other objects,
including an object from their survey and excluding HD 21362
\citep[due to free-free emission;][]{Rieke05}, had 24\,$\mu$m excesses
greater than 4 times their photospheric values at ages $>$20 Myr.
Implications of the IR excess fraction and values will be discussed in
$\S$5.\\

\begin{center}
 \begin{deluxetable*}{llrrrrllllc}
  \tabletypesize{\scriptsize}
  \tablewidth{0pt}
  \tablecaption{Stellar \& Disk Parameters}
  \tablehead{
  \colhead{ID} &
  \colhead{SpT.} &
  \colhead{$A_{V}$} &
  \colhead{$T_{*}$ (K)} &
  \colhead{log$\frac{L_{*}}{L_{\odot}}$} &
  \colhead{$\frac{M_{*}}{M_{\odot}}$} &
  \colhead{$\frac{L_{IR}}{L_{bol}}$} &
  \colhead{$T_{\rm disk}$ (K)} &
  \colhead{$D$ (AU)} &
  \colhead{Zodies} &
  \colhead{$L_{IR}>f_{max}$}
  }
  \startdata
32 Ori			&	B5V	&	0.06	&	15206	&	2.68	&	4.5	&\,\,\,\,\,	$\cdots$	&\,\,\,\,\,	$\cdots$	&\,\,\,\,\,	$\cdots$	&\,\,\,\,\,	$\cdots$	&	$\cdots$	\\
HD 35656$^{a}$		&	A0Vn	&	0.00	&	9707	&	1.37	&	2.2	&\,\,\,\,\,	1.5$\times10^{-4}$	&\,\,\,\,\,	94	&\,\,\,\,\,	42	&\,\,\,\,\,	2.6$\times10^{6}$	&	n	\\
HD 35714		&	A3	&	0.00	&	8503	&	1.06	&	1.8	&$<$	2.3$\times10^{-5}$	&$<$	22	&$<$	540	&$<$	6.3$\times10^{7}$	&	n	\\
HD 35499$^{b}$		&	F4	&	0.13	&	6609	&	0.43	&	1.3	&$<$	3.7$\times10^{-5}$	&$<$	215	&$<$	19	&$<$	1.2$\times10^{5}$	&	n	\\
			&		&		&		&		&		&$>$	3.0$\times10^{-5}$	&$>$	83	&$>$	3	&$>$	2.1$\times10^{3}$	&	n	\\
HD 36338$^{b}$		&	F4.5	&	0.00	&	6560	&	0.46	&	1.3	&$<$	1.4$\times10^{-4}$	&$<$	215	&$<$	14	&$<$	2.2$\times10^{5}$	&	n	\\
			&		&		&		&		&		&$>$	1.2$\times10^{-4}$	&$>$	97	&$>$	3	&$>$	1.0$\times10^{4}$	&	y	\\
HD 35695		&	F9	&	0.20	&	6042	&	0.25	&	1.2	&$<$	1.5$\times10^{-4}$	&$<$	52	&$<$	38	&$<$	2.0$\times10^{6}$	&	n	\\
HD 245567		&	G5	&	0.49	&	5738	&	0.23	&	1.2	&$<$	1.5$\times10^{-4}$	&$<$	50	&$<$	41	&$<$	2.3$\times10^{6}$	&	n	\\
HBC 443			&	G7	&	0.98	&	5560	&	0.28	&	1.3	&$<$	1.5$\times10^{-4}$	&$<$	58	&$<$	32	&$<$	1.5$\times10^{6}$	&	n	\\
RX\,J0520.5+0616$^{b}$	&	K3	&	0.60	&	4782	&	-0.35	&	0.9	&$<$	2.8$\times10^{-4}$	&$<$	215	&$<$	10	&$<$	2.8$\times10^{5}$	&	n	\\
			&		&		&		&		&		&$>$	1.4$\times10^{-4}$	&$>$	71	&$>$	1	&$>$	1.6$\times10^{3}$	&	y	\\
RX\,J0520.0+0612$^{a}$	&	K4	&	0.26	&	4600	&	-0.42	&	0.9	&\,\,\,\,\,	3.0$\times10^{-4}$	&\,\,\,\,\,	51	&\,\,\,\,\,	18	&\,\,\,\,\,	9.5$\times10^{5}$	&	n	\\
RX\,J0523.7+0652		&	K6.5	&	0.17	&	4164	&	-0.70	&	0.7	&$<$	6.5$\times10^{-4}$	&$<$	43	&$<$	18	&$<$	2.0$\times10^{6}$	&	n	\\
RX\,J0529.3+1210		&	K6.5	&	0.76	&	4164	&	-0.72	&	0.7	&$<$	7.5$\times10^{-3}$	&$<$	22	&$<$	63	&$<$	2.9$\times10^{8}$	&	n	\\
2UCAC 33699961		&	M3	&	0.09	&	3274	&	-1.30	&	0.2	&$<$	1.4$\times10^{-3}$	&$<$	51	&$<$	7	&$<$	6.0$\times10^{5}$	&	n	\\
2UCAC 33878694		&	M3	&	0.07	&	3274	&	-1.17	&	0.2	&$<$	6.3$\times10^{-3}$	&$<$	22	&$<$	41	&$<$	1.0$\times10^{8}$	& 
\enddata
\tablecomments{$^{a}$ Disk models constrained by two mid-IR data
  points, therefore values are not upper limits. $^{b}$ Disk models
  include both Wien temperature fit and modified blackbody fit, giving
  upper and lower limits for disk parameters. $^{c}$ Values of 1 and 0
  represent true and false, respectively, where $f_{max}$ is the
  maximum dust luminosity as defined by \citet{Wyatt07}. {\it Note to
    reader: These parameters will be superseded by those in Bell,
    Murphy, \& Mamajek (submitted to MNRAS).}}
\end{deluxetable*}
\end{center}

\section{Stellar and Disk Modeling \label{sec:diskmodel}}
\subsection{Stellar Photospheres}

We modeled spectral energy distributions (SEDs) of our target stars by
fitting ATLAS9 models of \citet{Castelli04} to $BVJHK_s$ \& IRAC
photometry.  For M-type stars (i.e. 2UCAC 33699961 \& 2UCAC 33878694),
we used NextGen model atmospheres \citep{Hauschildt99} since ATLAS9
models do not extend to that temperature range (i.e. $<$3500 K).  We
determined photospheric temperatures from spectral types and assumed
solar metallicity, surface gravity log$(g)\,=\,4.5$, and a
microturbulence of 2km/s for all objects.  We interpolated between
temperature bins of existing models to obtain new models corresponding
to the temperatures of our objects.  We computed model fluxes by
convolving atmospheric models with spectral response functions for
each passband (i.e.  Johnson $B$ \& $V$, 2MASS $JHK_s$, and IRAC
ch1-ch4).  The spectral response functions for the Johnson $B$ \& $V$
passbands were obtained from \citet{Bessel90}, 2MASS passbands were
obtained from the Asiago Database on Photometric Systems (ADPS)
\citep{Fiorucci03}, and the IRAC passbands for both full array and
subarray mode were obtained from the SSC
website\footnote{http://ssc.spitzer.caltech.edu/irac/calibrationfiles/spectralre-sponse/}.
We interpolated models to the wavelength range for each passband, then
determined model fluxes according to equation (3), where $F_{\lambda
  i}$ is the model flux at a given wavelength $\lambda_{i}$,
$T(\lambda_{i})$ is the spectral response function at that wavelength,
and $\triangle \lambda_{i}$ is the wavelength step.\\

\begin{equation}
F_{\lambda}=\frac{\sum{F_{\lambda i}T(\lambda_{i})\triangle \lambda_{i}}}{\sum{T(\lambda_{i})\triangle \lambda_{i}}}
\end{equation} 

ATLAS9 model atmospheres are sparsely sampled at wavelengths longer
than 20\,$\mu$m; therefore, we interpolated between points, adopting a
Rayleigh-Jeans functional dependence, as suggested in
\citet{Carpenter08} Appendix C2.  Additionally, we ``reddened'' model
fluxes according to the extinction curve defined by
\citet{Fitzpatrick07}, thereby allowing us to have $A_{V}$ as a
fitting parameter.  We used the mean Galactic curve determined for the
diffuse ISM and adopted R(V)\,=\,3.1.  Finally, we fit model fluxes to
observed fluxes by implementing bounded chi-squared minimization,
requiring that $A_{V}\geq0$.\\


\begin{figure}
 \epsscale{1.2} 
\plotone{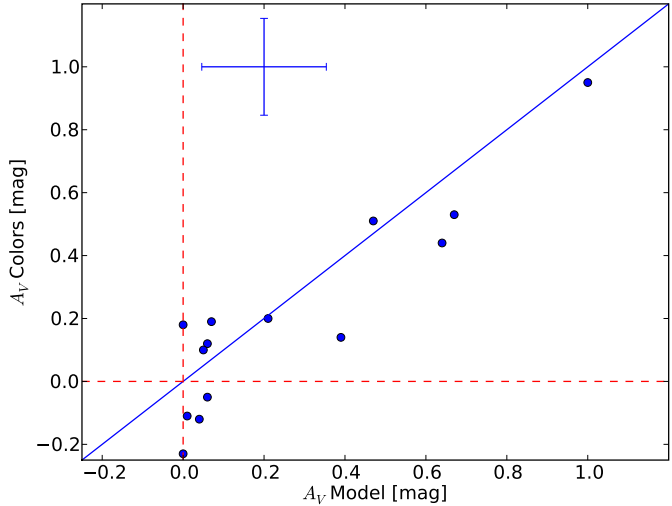} 
\caption{$A_{V,fit}$ vs. $A_{V, colors}$ for group members shows a
  near one-to-one correspondence.  Error bars correspond to an rms
  deviation of 0.15 mag.}
\end{figure}

Stellar parameters are presented in Table 2.  Final values for $A_{V}$
were computed by averaging $A_{V}$ derived from multiple colors (i.e.
$B-V$, $V-I_{c}$, $V-J$, $V-H$, $V-K_{s}$, $J-H$, $H-K_{s}$,
$J-K_{s}$, where available) with $A_{V}$ determined from the fitting
procedure, rounding negative values of $A_{V}$ to 0.  A plot of
$A_{V,fit}$ vs. $A_{V,colors}$ (Fig. 3) shows an approximate
one-to-one correspondence.  The rms of the scatter, 0.15 mag, was
comparable to the uncertainty in $A_{V}$ corresponding to a shift of
one spectral type.  Similarly, the rms was comparable to the standard
deviation of $A_{V}$ derived from multiple colors.  Thus, we conclude
that $A_{V}$ derived using both methods yield similar
values\footnote{{\it Updated estimates of the interstellar reddening
    in Bell, Murphy, and Mamajek (submitted to MNRAS) are now
    consistent with small values (E($B-V$) $\simeq$ 0.03\,$\pm$\,0.02
    mag) for both the hot and cool stars.}}.\\

\begin{figure}
 \epsscale{1.2}
 \plotone{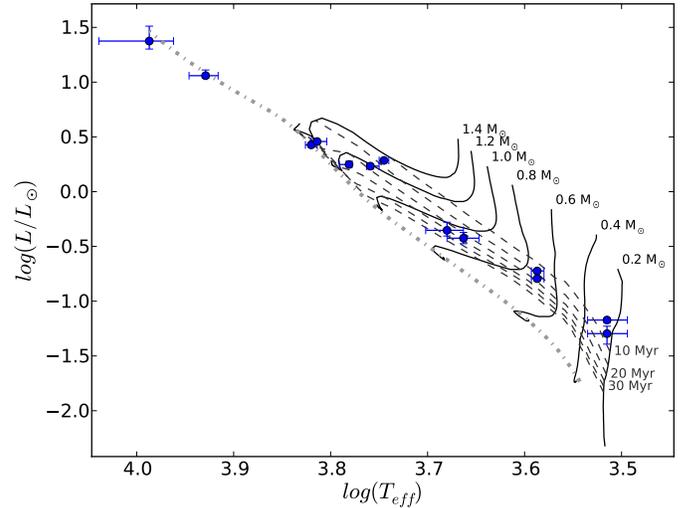}
 \caption{Theoretical H-R diagram for the observed 32 Ori members.
   Evolutionary tracks and isochrones are from \citet{Baraffe98},
   which do not extend to masses greater than 1.4 M$_{\odot}$.  We
   have also plotted a 250 Myr isochrone from \citet{Dotter08} to
   represent the main sequence for higher mass members. Error bars
   correspond to $\triangle L_{*}$ and $\triangle T_{*}$ resulting
   from a $\pm$ shift in spectral type.}
\label{hr_diagram}
\end{figure}

Stellar temperatures presented in Table 2 were derived from spectral
type and values for log$(L_{*}/L_{\odot})$ were obtained by
integrating unreddened model atmospheres.  Uncertainties in $T_{*}$
and log$(L_{*}/L_{\odot})$ correspond to a $\pm$ shift of one spectral
type.  Using these temperatures, luminosities, and their corresponding
uncertainties, we constructed a theoretical H-R diagram (Fig. 4).  We
interpolated the pre-main sequence evolutionary tracks of
\citet{Baraffe98} in order to fit the data and determined mean age for
the association of 21\,$\pm$\,15 Myr (1$\sigma$).  Stellar
photospheres are plotted with photometry in Figure 5.\\

\begin{figure*}
 \epsscale{1}
 \plotone{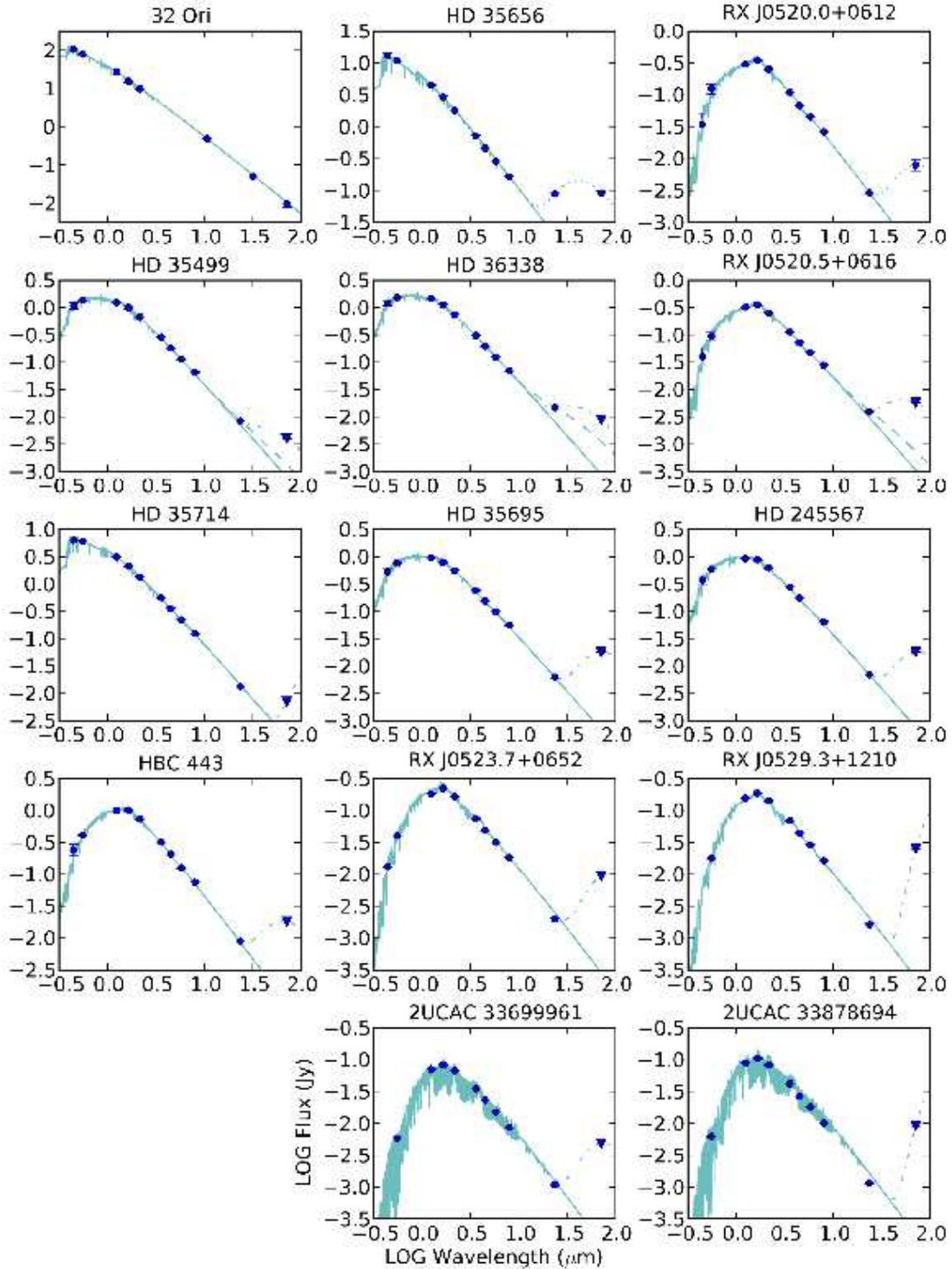}
 \caption{ATLAS9 model atmospheres were fit to $BVJHK_{s}$ \& IRAC
   data , when available (only $BVJHK_{s}$ data for 32 Ori).  NextGen
   model atmospheres were used for M-type stars (i.e. 2UCAC 33699961
   \& 2UCAC 33878694).  For HD 35656 and RX\,J0520.0+0612, modified
   blackbody curves were fit to both 24 \& 70\,$\mu$m detections
   (dotted line).  For the other 3 objects that exhibited 24\,$\mu$m
   excesses, upper and lower limits for disk parameters were obtained
   by fitting two curves: a blackbody curve corresponding to a
   24\,$\mu$m Wien temperature (dashed line), and modified blackbody
   curve fit to 24\,$\mu$m observations and 70\,$\mu$m 3$\sigma$ upper
   limits (dash-dotted line).  For all other objects, modified
   blackbody curves were used to model circumstellar disks by fitting
   24\,$\mu$m observations and 70\,$\mu$m 3$\sigma$ upper limits
   (dash-dotted curve).  These fits provide upper limits for disk
   parameters.}
\end{figure*}

\subsection{Dust Disks}

We sought to characterize mid-IR excesses of 32 Ori members by fitting
models of dust emission to the 24 and 70\,$\mu$m photometry.  We
modeled disk emission as either blackbody radiation or modified
blackbody radiation (i.e. $S_{\nu}\propto \nu^{\beta} B(\nu)$),
depending upon the particular circumstance, in order to obtain upper
and lower limits on disk parameters, like disk temperature,
luminosity, radius, and emitting area.  Lack of additional photometry
makes higher order disk modeling unwarranted as such models would be
poorly constrained.\\

Modeling disk emission as blackbody radiation gives an upper limit for
the true dust temperature \citep{Carpenter09}, whereas a modified
blackbody, which more closely resembles grain emission, yields a lower
limit for the dust temperature.  This is apparent from the Wien
displacement law, which yields a higher temperature for a simple
blackbody than for a modified blackbody (i.e. equation 4, which is
accurate to within 0.5\% for $\beta\in[0,2]$).\\

\begin{equation}
T_{BB}(\nu_{0})\approx(0.384\beta+1)\,T_{mod\,BB}(\nu_{0})
\end{equation}

\noindent Disk radius is inversely proportional to disk temperature in
that models that yield warmer disk temperatures predict smaller disk
radii; therefore, blackbody models give a lower limit on disk radius
and modified blackbody models give an upper limit.  The same is true
of the emitting area of the disk.  For the modified blackbody model
that we consider, we adopt $\beta\,=0.8\,$, following
\citet{Carpenter09}.  This value is derived from \citet{Williams06},
who constrained $\beta\in[0.6,1.0]$ with sensitive submillimeter
observations of a small sample of debris disks.\\

There were four permutations of 24 and 70\,$\mu$m excesses that were
observed: HD 35656 had both 24 and 70\,$\mu$m excesses; RX
J0520.0+0612 had no 24\,$\mu$m excess but a 70\,$\mu$m excess; HD
36338, RX\,J0520.5+0616, and HD 35499 had 24\,$\mu$m excesses and no
70\,$\mu$m detections; and the other eight objects had no 24\,$\mu$m
excesses and no 70\,$\mu$m detections.  32 Ori exhibited no mid-IR
excess; observations at 10.75, 32, and 70\,$\mu$m are all consistent
with photospheric emission.  We now consider these cases seperately.\\

First, we subtracted predicted photospheric emission from observed
fluxes and fit to these data.  For HD 35656 and RX\,J0520.0+0612, we
modeled the emission from the disk by fitting a modified blackbody
curve to the 24 and 70\,$\mu$m excess fluxes.  For HD 36338, RX
J0520.5+0616, and HD 35499, we obtained one range of limits by fitting
a blackbody curve to the 24\,$\mu$m excess flux only, adopting a
24\,$\mu$m Wien temperature.  Then we fit both the 24\,$\mu$m excess
flux and the 70\,$\mu$m 3$\sigma$ upper limit with a modified
blackbody curve.\\

Finally, for the 8 other objects that exhibited no 24\,$\mu$m
photospheric excess and whose 70\,$\mu$m flux was statistically zero,
we again fit modified blackbody curves to 24\,$\mu$m data and
70\,$\mu$m 3$\sigma$ upper limits, subtracting the photospheric
component.  Thus, these SEDs enabled us to compute upper limits for
disk parameters.  Values obtained from these fits are reported in
Table 2.\footnote{Parameters computed from 70\,$\mu$m 3$\sigma$ upper
  limits should be interpreted in the following way: there is a 99.7\%
  probability that the true parameter is less than the parameter that
  is presented.}\\

Ultimately, we determined dust temperatures within the range of
$T_{gr}\,=\,22-215$K and fractional infrared luminosities within the
range $L_{IR}/L_{*}\,=\,2.3\times10^{-5} - 6.3\times10^{-3}$, the
latter by integrating under the model.  As mentioned above, we
calculated two other important properties of the disk candidates that
we observed: the emitting area and grain distance.  Assuming that the
zodiacal emission from our own solar system is a single temperature
blackbody, with $T\,=\,260$\,K \citep{Reach96} and
$L_{IR}/L_{*}\,=\,8\times10^{-8}$ \citep[][see also \citet{Gaidos99}
  and \citet{Mamajek04}]{Good86}, the emitting area of dust around
another star in zodys ($Z$) is\\
 
\begin{equation}
A_{dust}=\left(\frac{T_{\rm disk}}{260 K}\right)^{-4}\left(\frac{L_{IR}/L_{*}}{8\times10^{-8}}\right) \,\, Z
\end{equation} 

\noindent Also, the ``characteristic grain distance'' D \citep{Chen05}, which is
derived from the radiative equilibrium equation of \citet{Jura98}, is\\

\begin{equation}
D=\frac{1}{2}\,\left(\frac{T_{*}}{T_{\rm disk}}\right)^{2}\,R_{*}
\end{equation}   

\noindent In this equation, $T_{*}$ is the stellar photospheric
temperature derived from spectral type, $T_{\rm disk}$ is the dust
grain temperature, and $R_{*}$ is the stellar radius derived from the
Stefan-Boltzmann equation
(i.e. $L_{*}\,=\,4\pi\,R^{2}_{*}\,\sigma\,T^{4}_{*}$).  These properties
are also listed in Table 2.\\

\begin{figure}[b]
 \epsscale{1.2} 
\plotone{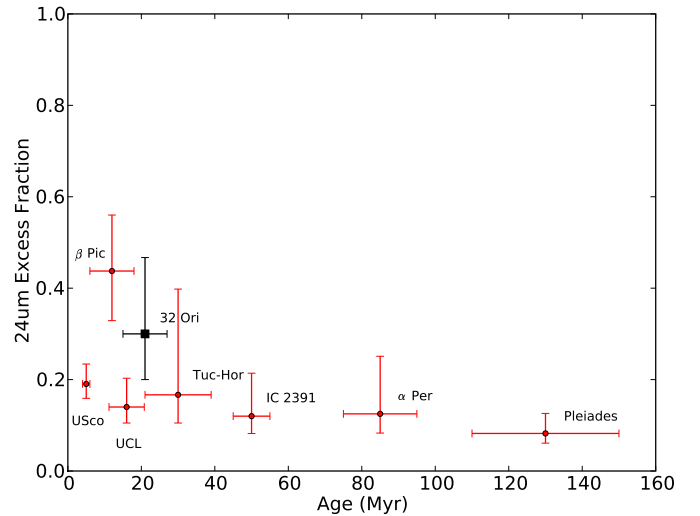} 
\caption{Debris disk fraction vs. age for the following groups: Upper
  Sco, $\beta$ Pic, UCL, Tucana-Horologium, IC 2391, $\alpha$\,Per,
  Pleiades (circles), and 32 Ori (square).  Error bars correspond to
  1$\sigma$ uncertainties in age and debris disk fraction, using
  binomial statistics.  Disk fraction, ages, and spectral type range
  were obtained from \citet{Carpenter09} and references therein.  {\it
    Note to reader: Updated ages from \citet{Pecaut16} (10\,$\pm$\,3
    Myr for US, 16\,$\pm$\,2 Myr for UCL) and \citet{Bell15}
    (24\,$\pm$\,3 Myr for $\beta$ Pic Moving Group and 45\,$\pm$\,4
    Myr for Tuc-Hor) supercede the values used in this figure from
    2010.}}
\end{figure}



\section{Discussion \label{sec:discussion}}

We place this group in context with previous studies of young stellar
associations by comparing its debris disk fraction with those of
groups compiled by \citet{Carpenter09}.  Several caveats must be taken
into consideration when making such comparisons.  Firstly, debris disk
fraction has been shown to depend on spectral type
\citep[e.g.][]{Carpenter09}.  Thus, in order not to be biased, it is
important that the debris disk surveys that we compare have comparable
sensitivity to lower mass stars; in other words, they cover the same
range in mass (i.e. spectral type).  We chose to include groups
encompassing a spectral type range from B-M, which is the range of
this study.  Secondly, as noted by \citet{Carpenter09}, various
surveys adopt different excess thresholds and it is therefore
necessary to define a common threshold.  Ultimately, for the seven
groups that we compare to 32 Ori (e.g. Upper Sco, $\beta$ Pic, UCL,
Tucana-Horologium, IC 2391, $\alpha$\,Per, and Pleiades), we used a
consistent 24\,$\mu$m excess threshold of $\geq$32\%.  A plot of
debris disk fraction vs. age for the groups that meet these standards
is presented in Figure 6.\\

We find that the debris disk fraction of the 32 Ori group is
consistent with disk fractions for groups of comparable ages and that
32 Ori is, in fact, the 2nd ``dustiest'' group after the $\beta$ Pic
moving group, among those lacking accreting stars.  The average excess
fraction of the seven groups is 18\% and the sample standard deviation
is 12\%.  Thus, the 24\,$\mu$m disk fraction of the 32 Ori group is
within 1$\sigma$ of this mean.\\

We now consider further the large excess exhibited by HD 35656
($\frac{L_{IR}}{L_{*}}=1.5\times10^{-4}$).  Is such an excess
representative of steady-state collisional evolution, or must it have
been caused by other means, such as a collisional event or feeding
from an outer disk?  \citet{Wyatt07} considered the collisional
evolution of debris disks around sun-like stars and derived a function
that defines the maximum dust luminosity of a disk as a function of
stellar mass, luminosity, age, and inner dust radius:
$L_{IR}/L_{\odot}<0.16\times10^{-3}M_{*}^{-5/6}L_{*}^{1/2}r_{\rm
  disk}^{7/3}t_{age}^{-1}$.  For a given object, excess emission above
this maximum luminosity would require an additional, ``transient''
production mechanism.  \citet{Wyatt07} note that the most likely
additional dust production mechanism is collisional grinding of
planetesimals scattered inward from an outer belt
\citep[e.g.][]{Beichman05}.  Although HD 35656 has among the largest
24\,$\mu$m excesses compared to stars $>$20 Myr, we found that its IR
luminosity was consistent with steady state disk evolution, according
to the equation of \citet{Wyatt07}.\\

Two stars, HD 36338 and RX\,J0520.5+0616, had IR luminosities that were
greater than their respective maximum dust luminosities, suggesting
that the IR excesses of these objects are unlikely to have resulted
from steady state evolution.  This result occurred only for models
that assumed a 24\,$\mu$m Wien temperature of $T\,=\,215\,K$.  Wien
temperature models produced IR luminosities that were nominally equal
to or less than models fit to both 24\,$\mu$m observations and
70\,$\mu$m 3$\sigma$ upper limits.  However, Wien temperature models
produced characteristic grain distances that were much smaller than
the latter model, resulting in a lower \textit{maximum} dust
luminosity.  If, in fact, these objects are simply undergoing steady
state evolution, their respective characteristic grain distances would
have to be $\geq$5 AU and $\geq$3 AU, assuming that their IR
luminosities remain the same.  Detection of IR excesses at longer
wavelengths would produce models with greater grain distances, likely
resulting in fractional IR luminosities that satisfy steady state disk
evolution.\\

\section{Summary \label{sec:summary}}

We have obtained IRAC 3.6, 4.5, 5.8 and 8\,$\mu$m photometry, MIPS 24
and 70\,$\mu$m photometry, and FAST low-resolution optical
spectroscopy photometry of 14 members of the 32 Ori group.  From the
optical spectroscopy, we determined spectral types and effective
temperatures for group members.  We chose ATLAS9 or NextGen models
corresponding to these temperatures and performed bounded chi-squared
fitting to optical and near-IR data.  By integrating over complete
atmospheric models, we determined more precise stellar luminosities
for group members, which enabled us to construct an HR diagram and
determine the age of the 32 Ori group using \citet{Baraffe98} pre-main
sequence tracks.  Using similar methods, we fit models of disk
emission to \textit{Spitzer} 24 and 70\,$\mu$m fluxes in order to
determine disk parameters.  Ultimately, we conclude the following:\\

\begin{itemize}

\item No 32 Ori members exhibit near-IR excesses in IRAC bands
  (3.6-8\,$\mu$m), and thus we conclude that they lack warm inner dust
  disks.  Spectroscopic results corroborate these findings and imply
  that none of the members are accreting.\\

\item 4 objects exhibit 24\,$\mu$m excesses at least 4$\sigma$ above
  their stellar photospheres, indicating that these members are debris
  disk candidates.  This corresponds to a 24\,$\mu$m disk fraction of
  29\% (+14\%, -9\%).  Only one of the 24\,$\mu$m excess objects, HD
  35656, had an excess at 70\,$\mu$m; however, we detected a
  70\,$\mu$m excess for RX\,J0520.0+0612, which did not have a
  24\,$\mu$m excess.\\

\item For the two objects which had both 24 and 70\,$\mu$m detections
  (HD 35656 and RX\,J0520.0+0612), their fractional infrared
  luminosities, disk temperatures, disk inner edges, and emitting
  areas are as follows:
  $L_{IR}/L_{*}=1.5\times10^{-4},\,3.2\times10^{-4}$;
  $T_{gr}\,=\,94,\,51\,$K; $D\,=\,43,\,18\,$AU; and
  $A_{dust}\,=\,2.6\times10^{6},\,9.5\times10^{5}\,Z$, respectively.
  Objects exhibiting 24\,$\mu$m excesses and no 70\,$\mu$m excess had
  disk parameters within the following ranges:
  $L_{IR}/L_{*}\,=\,3.0\times10^{-5} - 2.8\times10^{-4}$;
  $T_{gr}\,=\,71 - 215\,$K; $D\,=\,1 - 19\,$AU; and
  $A_{dust}\,=\,1.6\times10^{3} - 2.8\times10^{5}\,Z$.\\

\item The 32 Ori group is similar to groups with comparable ages in
  its debris disk fraction and has the second highest 24\,$\mu$m
  excess fraction among groups lacking accreting T Tauri stars (behind
  only the $\beta$ Pic moving group).  Examining the fractions of
  debris disks, as determined by exhibiting a 24\,$\mu$m excess
  $\geq$32\% of the stellar photosphere, within groups encompassing a
  mass range from B-M spectral type, we find that the disk fraction
  within the 32 Ori group is statistically consistent with the average
  disk fraction of the 7 groups that fit this criteria.\\

\item HD 35656 has a 24\,$\mu$m excess that is among the highest for
  objects $>$20 Myr; however, its IR luminosity is consistent with
  steady state disk evolution.  Wien temperature models of HD 36338
  and RX\,J0520.5+0616 produce IR luminosities that exceed their
  respective maximum dust luminosities.  A transient event may be
  required to explain such high luminosities; however, a larger
  characteristic grain distance than predicted by the Wien model would
  be consistent with steady state disk evolution.\\

\end{itemize}

The characteristics of the debris disks in the 32 Ori group in
addition to the overall frequency of disks within the group seem
typical compared to groups of similar ages.  If more low mass 32 Ori
group members exist, perhaps their statistics will yield a lower
overall debris disk fraction.  The proximity of the group makes 32 Ori
a unique laboratory to further examine debris disk properties.\\

\acknowledgements

This work is based on observations made with the \textit{Spitzer Space
  Telescope}, which is operated by JPL/Caltech under a contract with
NASA.
The project was supported by NASA contract 1361947, administered
through JPL.
Part of this research was carried out at the Jet Propulsion
Laboratory, California Institute of Technology, under a contract with
the National Aeronautics and Space Administration.
EEM and MJP also acknowledge support from the School of Arts and
Sciences from the University of Rochester.
This publication makes use of data products from the Two Micron All
Sky Survey, which is a joint project of the University of
Massachusetts and the Infrared Processing and Analysis
Center/California Institute of Technology, funded by the National
Aeronautics and Space Administration and the National Science
Foundation.
This paper has been approved for unlimited release (record URS263148,
CL \#16-6146).
Facilities: Spitzer, CTIO:1.5m, CTIO:2MASS, FLWO:2MASS, FLWO:1.5m,
HIPPARCOS.\\





\end{document}